\documentclass[aps,preprint,groupedaddress,showkeys]{revtex4-1}
\usepackage{float}
\usepackage{multirow}
\usepackage{amsmath}
\usepackage{amssymb}
\usepackage{graphicx}
\usepackage{stackrel}
\usepackage{natbib}
\usepackage[unicode=true]{hyperref}

\begin{document}
	\title{Bounds on abundance of primordial black hole and dark matter from EDGES 21-cm signal}
	
	\author{Ashadul Halder}
	\email{ashadul.halder@gmail.com}
	\affiliation{Department of Physics, St. Xavier's College, \\30, Mother Teresa Sarani, Kolkata-700016, India.}
	
	\author{Shibaji Banerjee}
	\email{shiva@sxccal.edu}
	\affiliation{Department of Physics, St. Xavier's College, \\30, Mother Teresa Sarani, Kolkata-700016, India.}
	
	\date{\today}
	\begin{abstract}
		The redshifted 21cm radio signal has emerged as an important probe for investigating the dynamics of the dark age Universe (recombination to reionization). In the current analysis, we explore the combined effect of primordial black hole (PBH) evaporation and the baryon-dark matter (DM) interaction in the 21cm scenario. The variation of brightness temperature shows remarkable dependence on the DM masses ($m_{\chi}$) and the baryon-DM cross-sections ($\overline{\sigma}_0$) besides the influences of the PBH parameters (mass $\mathcal{M_{\rm BH}}$ and initial mass fraction $\beta_{\rm BH}$). We address both upper and lower bounds on $\beta_{\rm BH}$ for a wide range of PBH mass in presence of different $m_{\chi}$ and $\overline{\sigma}_0$ by incorporating the observational excess $\left(-500^{+200}_{-500}\: {\rm mK}\right)$ of EDGES's experimental results. Finally, we address similar limits in the $m_{\chi}$ - $\overline{\sigma}_0$ parameter plane for different PBH masses.

	\end{abstract}
	\maketitle

\section{Introduction}
	The dynamic of the Universe in the dark age is still unexplored due to the lack of luminous sources. The 21cm neutral hydrogen spectrum can be a promising probe in understanding the dynamics of the early Universe particularly during this unexplored era. The redshifted signature of the 21cm hydrogen absorption spectrum may provide a detailed understanding regarding the reionization and the Primordial Black Holes (PBHs) \cite{BH_21cm_0,BH_21cm_1,BH_21cm_2,BH_21cm_4,BH_21cm_5,legal_1} as well as the baryon-dark matter (DM) scattering and neutrino physics \cite{21cm_nu_1,21cm_nu_1} in the high redshifted epoch.
	
	Hydrogen is the most abundant baryonic component of the Universe, which occupies $\sim 75\%$ of the entire baryonic budget. The 21cm ($\sim 1.42$ GHz) hyperfine spectrum manifests as an outcome of the transition between the electronic spin states (s=0,1) of Hydrogen atoms. The corresponding brightness temperature $T_{21}$ represents the intensity of the spectrum which mainly depends on the quantity $T_s - T_{\gamma}$. Here, $T_{\gamma}$ ($T_{\gamma}=2.725 (1+z)$ K) and $T_s$ are the cosmic microwave background (CMB) temperature and the spin temperature respectively. The spin temperature indicates the population of the hydrogen atom with different energy states. The ``Experiment to Detect the Global Epoch of Reionization Signature'' (EDGES) \cite{edges} reported a prominent footprint of 21cm line (21cm brightness temperature $-500^{+200}_{-500}$ mK) at cosmic dawn ($14<z<20$) with $99\%$ confidence level (C.L.). But according to the notion of the standard cosmology, the estimated brightness temperature is only $\approxeq-200$ mK.
	As a consequence, the additional cooling observed by the EDGES experiment can be explained either by enhancing the background temperature $T_{\gamma}$ or by lowering the baryonic temperature which is almost equal to the spin temperature at $14<z<20$. The evaporation of primordial black holes (PBHs) can be such a possible source that heats up the intergalactic medium (IGM) resulting in the rise in background temperature. Dark matter annihilation, decay, even the baryon-dark matter interaction are also the possible sources that can induce the larger than the expected separation between $T_s$ and $T_{\gamma}$ \cite{rdi,legal_2,atri}. In the present work, we investigate the combined effect of the PBH evaporation and the baryon-DM interaction in the framework of the global 21cm signature.
	
	The idea of primordial Black Hole (PBH) was first introduced by \citet{PBH_0}. The PBHs are believed to be originated due to the collapse of the overdensity region during the early epoch of the Universe \cite{khlopov_1,khlopov_2,khlopov_3,juan}. The limit of the overdensity is characterized by the Jeans length $R_j$, $R_j = \sqrt{\displaystyle\frac {1} {3G \rho}}$. The density fluctuation $\delta$ also has to satisfy the condition $\delta_{\rm min} \leq \delta \leq \delta_{\rm max}$ ($\delta_{\rm max}$ and $\delta_{\rm min}$ are the threshold density contrasts). The fluctuation $\delta \rho$ is given by $\rho = \rho_c + \delta \rho$, where $\rho_c$ is the critical collapse density. The gravitational fluctuation during the inflation is the most acceptable conjecture of the PBH formation \cite{stpbh_1,stpbh_2,stpbh_3,stpbh_4,stpbh_5,stpbh_6,stpbh_7}. Besides the standard scenarios, there are several alternative mechanisms describing the formation of PBHs namely, fragmentation of scalar condensation \cite{fsc_1,fsc_2,fsc_3}, collapse of cosmic strings and domain walls \cite{ccs_1,ccs_2,ccs_3} etc. It is suggested that, smaller PBHs may evaporate entirely very rapidly, a PBH of mass $\geq 10^{14}$ g has the lifetime of the order of the cosmological timescale. In the work of \citet{BH_21cm_1}, the authors investigated the global 21cm signal due to PBH evaporation having masses $\geq 10^{15}$ g. On the other hand, in the work of \citet{BH_21cm_2}, the author focused on the PBH mass $\lesssim 10^{14}$ g. In the current analysis, we look for the 21cm signal for a wide range of PBH masses $10^{13} \leq \mathcal{M}_{\rm BH}\leq 10^{15}$ g and hence the corresponding bounds on initial mass fraction of PBHs for different DM masses.
	
	We tried to expose the effect of baryon-DM interaction along with the heating due to the PBHs in the evolution of brightness temperature. Throughout the calculation, the baryon-DM interaction cross-section ($\bar{\sigma}$) is parameterized as $\bar{\sigma}=\sigma_0 v^{-4}$ \cite{munoz}, where the considered dark matter is assumed to be model-independent \cite{IDM_1,IDM_2,Cheng:2002ej,servant_tait,Hooper:2007gi,Majumdar:2003dj,wimpfimp}. In the case of PBH heating, only the effect of the Hawking radiation is taken into account. The current analysis is mainly focused on the investigation of the bounds on PBH and DM parameters and their mutual influences with other parameters. We address both the upper and lower bound of the initial mass fraction of PBHs ($\beta_{\rm BH}$) for a wide range of PBH mass and compare the upper bound with the same as obtained from the work of \citet{BH_21cm_3,BH_21cm_2}. We also compare our result with the limits, obtained by evaluating the \texttt{COSMOMC} code with the Planck-2015 data \cite{planck15,BH_21cm_3} and 21cm power spectra \cite{BH_21cm_2,BH_21cm_3}. Besides the PBH parameters, the bound on $\sigma_{41}$ and $m_{\chi}$ and their variations are also described using the demonstrative plots. The limits obtained from this particular analysis agree with the work of \citet{rennan_3GeV}.
	
	The paper is organized as follows. Section~\ref{sec:PBH} deals with the energy injection by the PBHs in the form of Hawking radiation. In Section~\ref{sec:T_evol} and Section~\ref{sec:21cm}, the thermal evolutions are described in presence of baryon-DM interaction and PBHs. The results and corresponding plots are furnished in Section~\ref{sec:result}. Finally in Section~\ref{sec:conc}, concluding remarks are given.

\section{\label{sec:PBH} The Influences of Primordial Black Holes}
	The low mass black holes can be a possible source of IGM heating and hence the 21cm signal. The radiation of steady $e^{\pm}$ and $\gamma$ in the form of Hawking Radiation \cite{BH_F} can modify the global 21cm brightness temperature significantly. Several recent works and numerical simulations \cite{BH_21cm_0,BH_21cm_1,BH_21cm_2,BH_21cm_3,BH_21cm_4,BH_21cm_5,amar21} verify the effect of PBHs in this context. Moreover, in the work of \citet{BH_21cm_1}, one can see that, the IGM heating by the Hawking radiation and the same due to the dark matter decay are equally significant in global 21cm signature.
	
	A black hole of mass $M_{\rm{BH}}$ evaporates at the rate \cite{hawking,BH_21cm_1}
	\begin{equation}
		\dfrac{{\rm d}M_{\rm{BH}}}{{\rm d}t} \approx -5.34\times10^{25} \left(\sum_{i} \mathcal{F}_i\right) \left(\dfrac{M_{\rm{BH}}}{\rm g}\right)^{-2} \,\,\rm{g/sec}
		\label{eq:PBH}
	\end{equation}
	where, the coefficient $\mathcal{F}_i$ represents the fraction of evaporation in the form of $i^{\rm th}$ particle. The evaporation fraction $\sum_{i} \mathcal{F}_i$ mainly depends on the PBH temperature $T_{BH}$, given by \cite{BH_F},
	\begin{eqnarray}
		\sum_{i} \mathcal{F}_i&=&1.569+0.569 \exp 
		\left(-\frac{0.0234}{T_{\rm{BH}}} \right)+
		3.414\exp \left(-\frac{0.066}{T_{\rm{BH}}} \right)\nonumber\\
		&&+ 1.707\exp \left(-\frac{0.11}{T_{\rm{BH}}} \right)+ 
		0.569\exp \left(-\frac{0.394}{T_{\rm{BH}}} \right)\nonumber\\
		&&+1.707\exp \left(-\frac{0.413}{T_{\rm{BH}}} \right)+ 
		1.707\exp \left(-\frac{1.17}{T_{\rm{BH}}} \right)\nonumber\\
		&&+1.707\exp \left(-\frac{22}{T_{\rm{BH}}} \right)+
		0.963\exp \left(-\frac{0.1}{T_{\rm{BH}}}\right)
	\end{eqnarray}
	The PBH temperature $T_{\rm BH}$ can be estimated from the PBH mass from the relation, $T_{\rm{BH}}=1.05753 \times \left(M_{\rm{BH}}/10^{13} {\rm g}\right)^{-1}$ GeV. The rate of energy injected in the form of Hawking radiation is described as \cite{BH_21cm_2,amar21},
	\begin{equation}
		\left.\dfrac{{\rm d} E}{{\rm d}V {\rm d}t}\right|_{\rm{BH}}=\dfrac{1}{M_{\rm{BH}}}\dfrac{{\rm d} M_{\rm{BH}}}{{\rm d} t} n_{\rm BH}(z)
	\end{equation}
	where, $n_{\rm{BH}}(z)$ is the PBH number density at redshift $z$, given by \cite{BH_21cm_2},
	\begin{eqnarray}
		n_{\rm{BH}}(z)&=&\beta_{\rm BH}\left(\dfrac{1+z}{1+z_{\rm eq}}\right)^3 \dfrac{\rho_{\rm c,eq}}{\mathcal{M}_{{\rm BH}}} \left(\dfrac{\mathcal{M}_{\rm H,eq}}{\mathcal{M}_{\rm H}}\right)^{1/2} \left(\dfrac{g^i_{\star}}{g^{\rm eq}_{\star}}\right)^{1/12}\nonumber\\
		&\approx&1.46 \times 10^{-4}\beta_{\rm BH} \left(1+z\right)^3 \left(\dfrac{\mathcal{M}_{{\rm BH}}}{\rm g}\right)^{-3/2} {\rm cm^{-3}}
	\end{eqnarray}
	In the above expression, $\mathcal{M_{\rm BH}}$ is the mass of the PBH at the time of formation and $\mathcal{M_{\rm H}}$ is the horizon mass \cite{BH_21cm_2,BH_21cm_3,betabh}. The quantity $\beta_{\rm BH}$ represents the initial mass fraction of PBHs.
	
\section{\label{sec:T_evol} Temperature Evaluation of IGM}
	In our analysis, the thermal evolution of the charge-neutral Universe ($x_e = x_p$, where $x_e$ and $x_p$ are the abundances of the electron and proton respectively) is studied by evolving the dark matter temperature ($T_{\chi}$) and the baryon temperature ($T_b$) with cosmological redshift $z$. After incorporating the effects of energy injection from PBH evaporation and the baryon-DM interaction, the evolution equations ($T_{\chi}$ and $T_{b}$ of Ref.~\cite{munoz}) take the form \cite{corr_equs,munoz,BH_21cm_1,BH_21cm_2,amar21},
	\begin{equation}
		(1+z)\frac{{\rm d} T_\chi}{{\rm d} z} = 2 T_\chi - \frac{2 \dot{Q}_\chi}{3 H(z)}, 
		\label{eq:T_chi}
	\end{equation}
	\begin{equation}
		(1+z)\frac{{\rm d} T_b}{{\rm d} z} = 2 T_b + \frac{\Gamma_c}{H(z)}
		(T_b - T_{\gamma})-\frac{2 \dot{Q}_b}{ 3 H(z)}-\frac{2}{3 k_B H(z)} \frac{K_{\rm BH}}{1+f_{\rm He}+x_e}, 
		\label{eq:T_b}
	\end{equation}
	In Eq.~\ref{eq:T_b}, the last term 
	appears due to the energy deposition in the form of Hawking radiation \cite{BH_21cm_1,BH_21cm_2,amar21}. In the above expression (Eq.~\ref{eq:T_b}), $\Gamma_c$ describes the effect of the Compton scattering ($\Gamma_c=\frac{8\sigma_T a_r T^4_{\gamma}x_e}{3(1+f_{\rm He}+x_e)m_e c}$) where $\sigma_r$ and $a_T$ are the radiation constant and the Thomson scattering cross-section respectively. $x_e$ and $f_{\rm He}$ are the fractional abundance of electron and He respectively. In the current calculation, the heating rates of the baryonic fluid $\dot{Q}_b$ and DM fluid $\dot{Q}_{\chi}$ due to the baryon-DM interaction are estimated according to the work of \citet{munoz}. The fluid terms ($\dot{Q}_b$ and $\dot{Q}_{\chi}$) mainly depend on the drag term $(V_{\chi b}\equiv V_{\chi}-V_{b})$ ($V_{\chi}$ and $V_{b}$ are the velocity terms for the dark matter and baryon respectively). The evolution equation of the drag term is defined as \cite{munoz},
	\begin{equation}
		\frac{{\rm d} V_{\chi b}}{{\rm d} z} = \frac{V_{\chi b}}{1+z}+
		\frac{D(V_{\chi b})}{(1+z) H(z)}, \label{eq:V_chib}
	\end{equation}
	where $D(V_{\chi b})$ is expressed as, 
	\begin{equation}
		D(V_{\chi b})=\dfrac{{\rm d}(V_{\chi b})}{{\rm d}t}=\dfrac{\rho_m 
			\sigma_0}{m_b + m_{\chi}} \dfrac{1}{V^2_{\chi b}} F(r).
		\label{eq:dvchib}
	\end{equation}
	In the above equation, $F(r)={\rm erf}\left(r/\sqrt{2}\right)-\sqrt{2/\pi}r e^{-r^2/2}$ ($\rm erf$ represents the error function), $r=V_{\chi b}/u_{\rm th}$ and $u_{\rm th}^2=T_b/m_b+T_{\chi}/m_{\chi}$ is the variance of the relative thermal motion. The scattering cross-section parameter $\sigma_0$ is parametrized as a dimensionless quantity $\sigma_{41}=\frac{\sigma_0}{10^{-41} {\rm cm^2}}$.
	
	Along with $T_{\chi}$ and $T_b$, the free electron abundance $x_e$ of the IGM is also perturbed remarkably due to the energy deposition from PBHs. THe electron abundance $x_e$ depends on $T_b$ and $T_{\gamma}$ simultaneously given by \cite{munoz, BH_21cm_2, amar21},	
	\begin{equation}
		\frac{{\rm d} x_e}{{\rm d} z} = \frac{1}{(1+z)\,H(z)}\left[I_{\rm Re}(z)-
		I_{\rm Ion}(z)-I_{\rm BH}(z)\right], 
		\label{eq:xe}
	\end{equation}
	In the above expression, $I_{\rm Re}(z)$ and $I_{\rm Ion}(z)$ describe the standard recombination and ionization rate respectively, given by \cite{yacine,hyrec11,munoz},
	\begin{equation}
		I_{\rm Re}(z)-I_{\rm Ion}(z) = C_P\left(n_H \alpha_B x_e^2-4(1-x_e)\beta_B 
		e^{-\frac{3 E_0}{4 k_B T_{\gamma}}}\right),
		\label{eq:xe_comp}
	\end{equation}
	where $\alpha_B$ and $\beta_B$ represent the case B recombination coefficient and the photoionization coefficient respectively and $C_P$ is the Peebles C factor \cite{peeble,hyrec11}. 
	
	The case B recombination coefficient ($\alpha_B$) (in $\rm{m^3}s^{-1}$) can be estimated by fitting the data obtained in the work of \citet{pequignot} as \cite{pequignot,BH_21cm_5},
	\begin{equation}
		\alpha_B=10^{-19} F\left(\frac{a t^b}{1+c t^d}\right), \label{alphaB}
	\end{equation}
	where, $t$ represents the temperature in $10^4$K \cite{hummer,pequignot,seager} and the fitted parameters are $a=4.309$, $b=-0.6166$, $c=0.6703$, $d=0.5300$, $F=1.14$ \cite{pequignot,BH_21cm_5}. The photoionization rate ($\beta_B$) is given by \cite{seager,BH_21cm_5},
	\begin{equation}
		\beta_B=\alpha_B \left(\frac{2 \pi \mu_e k_B T_{\gamma}}{h^2}\right)^{3/2} 
		\exp\left(-\frac{h \nu_{2s}}{k_B T_{\gamma}}\right), \label{betaB}
	\end{equation}
	where, $\nu_{2s}$ is the frequency of the emitted photon during the $2s\rightarrow 1s$ transition and $\mu_e$ is the reduced mass of the neutral hydrogen atom. The Peebles C factor is described in terms of escape rate of the ${\rm Ly\alpha}$ photons ($R_{\rm Ly\alpha}=8\pi H/\left(3 n_H (1-x_e)\lambda_{\rm Ly\alpha}^3\right)$) as \cite{hyrec11,peeble},
	\begin{equation}
		C_P=\dfrac{\frac{3}{4}R_{\rm Ly\alpha}+\frac{1}{4}\Lambda_{2s1s}}{\beta_B+
			\frac{3}{4}R_{\rm Ly\alpha}+\frac{1}{4}\Lambda_{2s,1s}}, \label{peeblec}
	\end{equation}		
	where $\Lambda_{2s,1s}\approx 8.22 \, \rm{s^{-1}}$ \cite{hyrec11}. 
	
	The term $I_{\rm BH}$ appears in the expression of electron abundance evolution (Eq.~\ref{eq:xe}) is described as
	\begin{equation}
	 I_{\rm BH}=\chi_i f(z) \frac{1}{n_b} \frac{1}{E_0}\times \left.
	 \dfrac{{\rm d} E}{{\rm d}V {\rm d}t}\right|_{\rm{BH}}. \label{IBH}
	\end{equation}
	In the evolution equation of the baryon temperature (Eq.~\ref{eq:T_b}), the term $K_{\rm BH}$ arises due to PBH evaporation and is given by,
	\begin{equation}
		K_{\rm BH}=\chi_h f(z) \frac{1}{n_b} \times \left.\dfrac{{\rm d} E}{{\rm d}V {\rm d}t}\right|_{\rm{BH}}. \label{KBH}
	\end{equation}
	where $E_0=13.6$ eV, $\chi_i=(1-x_e)/3$ and $\chi_h=(1+2x_e)/3$ are the fraction of the deposited energy, that contributes in the IGM ionization and heating respectively \cite{BH_21cm_2,chen,BH_21cm_4,PhysRevD.76.061301,Furlanetto:2006wp}. In Eqs.~\ref{IBH} and \ref{KBH}, the parameter $f(z)$ is the ratio of total amount of energy deposited to the injected energy \cite{corr_equs,fcz001,fcz002,fcz003,fcz004}.
		
\section{\label{sec:21cm} 21cm Signal}
	The neutral hydrogen atom has two electronic hyperfine spin states (spin 0 and spin 1). The 21cm absorption line of the hydrogen atom is originated due to the transition of electron between those two hyperfine states, and described by the brightness temperature $T_{21}$. The brightness temperature ($T_{21}$) basically represents the intensity of the 21cm line at different values of cosmological redshift $z$. The expression of brightness temperature is given by,
	\begin{equation}
		T_{21}=\dfrac{T_s-T_{\gamma}}{1+z}\left(1-e^{-\tau(z)}\right)
		\label{eq:t21}
	\end{equation}
	where, $T_s$ is the 21cm spin temperature at redshift $z$ and $\tau(z)$ is the optical depth of the IGM, given by \cite{munoz},
	\begin{equation}
		\tau(z) = \dfrac{3}{32 \pi}\dfrac{T_{\star}}{T_s}n_{\rm HI} \lambda_{21}^3\dfrac{A_{10}}{H(z)+(1+z)\delta_r v_r}.
		\label{eq:tau}
	\end{equation}
	In the above expression $T_{\star}$ ($=hc/k_B \lambda_{21}=0.068$ K), $A_{10}=2.85\times 10^{-15}\,{\rm s^{-1}}$ is the Einstein coefficient \cite{yacine,hyrec11}, $\lambda_{21}\approx 21$ cm and $\delta_r v_r$ is the radial gradient of the peculiar velocity.
	
	The spin temperature ($T_s$) is characterized by, 
	\begin{equation}
		\frac{n_1}{n_0}=3 \exp{-\frac{T_{\star}}{T_s}},
	\end{equation}
    where, $n_1$ and $n_0$ are the number density of neutral hydrogen atoms at excited and ground state respectively. The resonant scattering of Ly$\alpha$ photons and the background photons mainly modify the spin temperature. On the other hand, the heating in the form of Hawking radiation and baryon-DM scattering also perturb the $T_s$ remarkably (see Fig.~\ref{fig:tspin}). At equilibrium, the expression of $T_s$ can be expressed as,
	\begin{equation}
		T_s = \dfrac{T_{\gamma}+y_c T_b+y_{\rm Ly\alpha} T_{\rm Ly\alpha}}
		{1+y_c+y_{\rm Ly\alpha}},
		\label{eq:tspin}
	\end{equation}
	where, the term $y_{\rm LY\alpha}$ arises due to the Wouthuysen-Field effect. $T_{\rm Ly\alpha}$ and $y_c$ are the Lyman-$\alpha$ background temperature and the collisional coupling parameter respectively \cite{BH_21cm_1}. The coefficients $y_c$ and $y_{\rm LY\alpha}$ are given by $y_c=\frac{C_{10}T_{\star}}{A_{10} T_b}$ and $y_{\rm LY\alpha}=\frac{P_{10}T_{\star}}{A_{10} T_{\rm Ly\alpha}}e^{0.3 \times (1+z)^{1/2} T_b^{-2/3} \left(1+\frac{0.4}{T_b}\right)^{-1}}$ \cite{BH_21cm_2,Yuan_2010,Kuhlen_2006}. Here, $P_{10}$ is the deexcitation rate due to Lyman-$\alpha$ given by $P_{10}\approx1.3\times 10^{-21}S_{\alpha}J_{-21}\,{\rm s^{-1}}$ where $C_{10}$ is the collision deexcitation rate. $S_{\alpha}$ is the spectral distraction factor \cite{salpha} and $J_{-21}$ represents the Lyman-$\alpha$ background intensity \cite{jalpha} respectively.
	
\section{\label{sec:result} Calculation and Results}
	In the current work, we looked into the 21cm signature due to Hawking radiation, in the presence of the effect of the baryon-DM interaction. In order to investigate the variation of the spin temperature ($T_s$) and hence the brightness temperature ($T_{21}$), five mutually coupled equations are evolved (Eqs.~\ref{eq:PBH}, \ref{eq:T_chi}, \ref{eq:T_b}, \ref{eq:xe} and \ref{eq:V_chib}) together with redshift $z$.	
	\begin{figure}
		\includegraphics[width=0.7\linewidth]{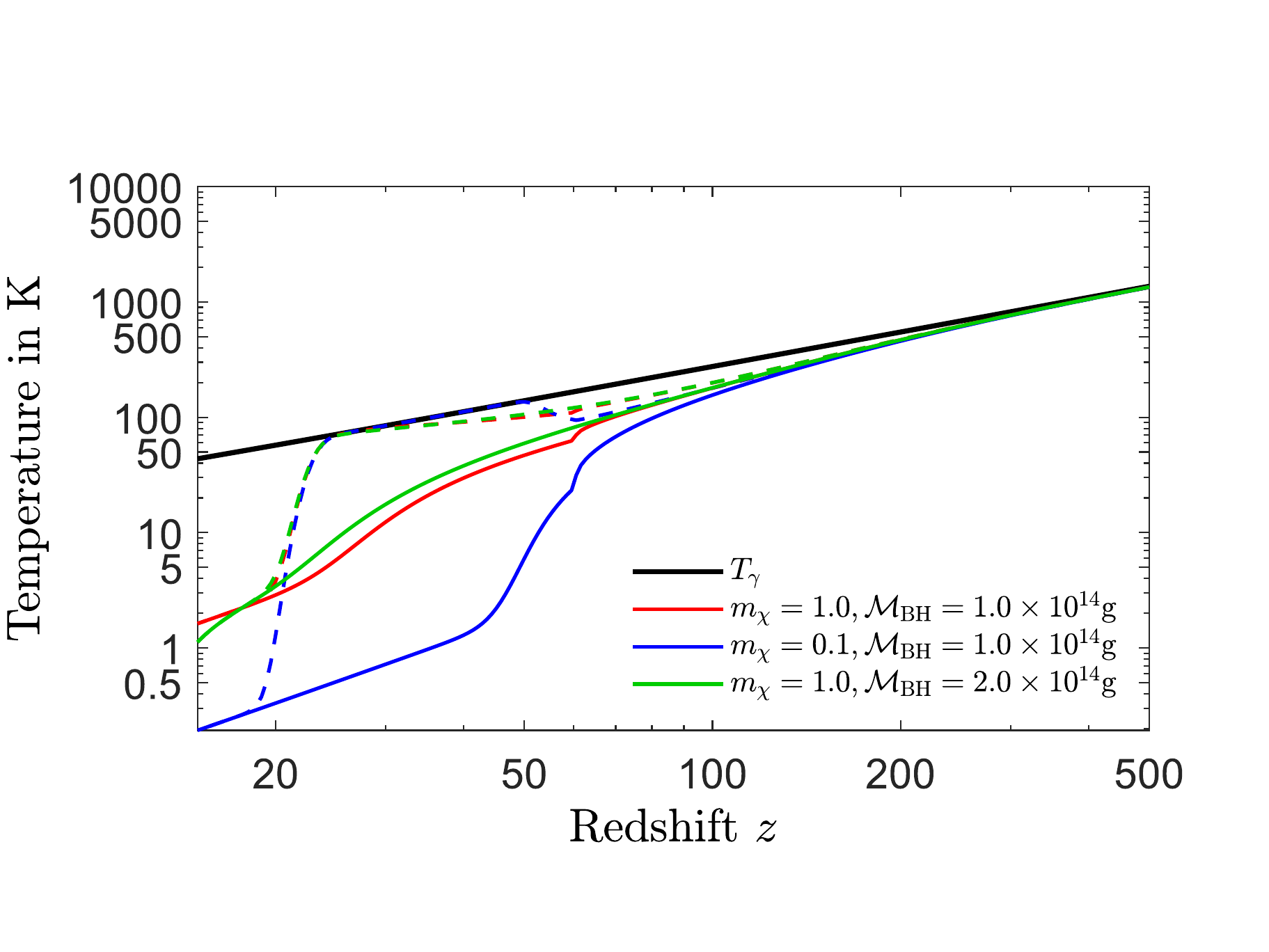}
		\caption{\label{fig:tspin} Evolution of $T_b$ (solid line) and $T_s$ (dashed line) with cosmological redshift ($z$) for different sets of DM mass $m_{\chi}$ and PBH mass $\mathcal{M_{\rm BH}}$. $T_{\gamma}$ is represented by the solid black line. In each case, $\sigma_{41}=1$ and the $\beta_{\rm BH}=10^{-29}$ are considered.}
	\end{figure}
	In Fig.~\ref{fig:tspin}, the variation of $T_b$ and $T_s$ with $z$ is described for different values of PBH masses and DM masses. The solid red line shows the evolution of $T_b$ where the chosen mass of the DM and the PBHs is $m_{\chi}=1$~GeV $\mathcal{M}_{\rm BH}=10^{14}$ g respectively. The spin temperature for the corresponding case is represented by the dashed red line. The blue and the green solid lines are indicating the baryon temperature for the sets $m_{\chi}=0.1$~GeV, $\mathcal{M}_{\rm BH}=10^{14}$ g and $m_{\chi}=1$~GeV, $\mathcal{M}_{\rm BH}=2 \times 10^{14}$ g respectively. The dashed blue and green lines are representing the corresponding spin temperatures. For each of the cases described in Fig.~\ref{fig:tspin}, we choose $\sigma_{41}=1$ and $\beta_{\rm BH}=10^{-29}$.
		
	\begin{figure*}
		\centering{}
		\begin{tabular}{cc}
			\includegraphics[width=0.48\textwidth]{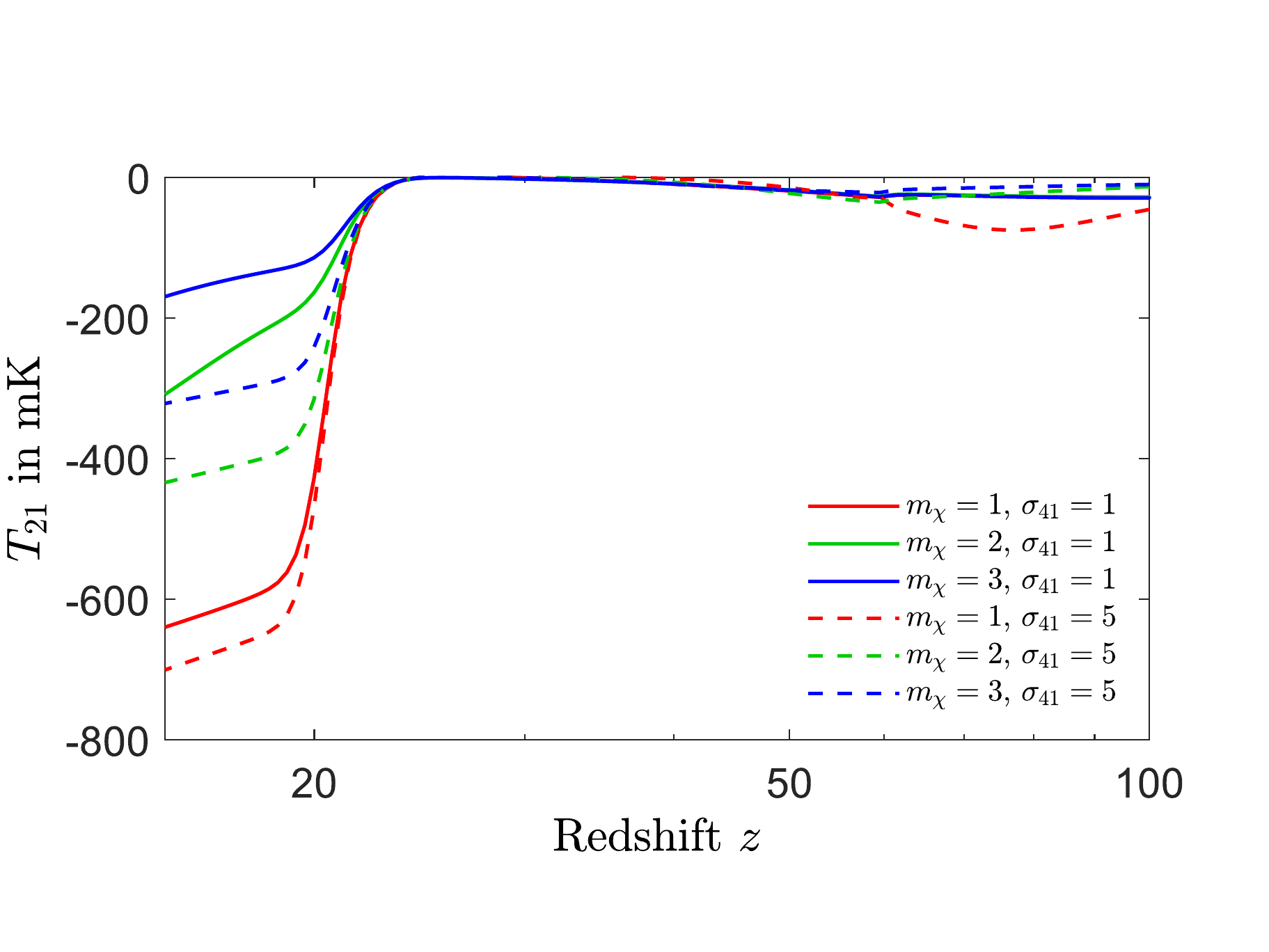}&
			\includegraphics[width=0.48\textwidth]{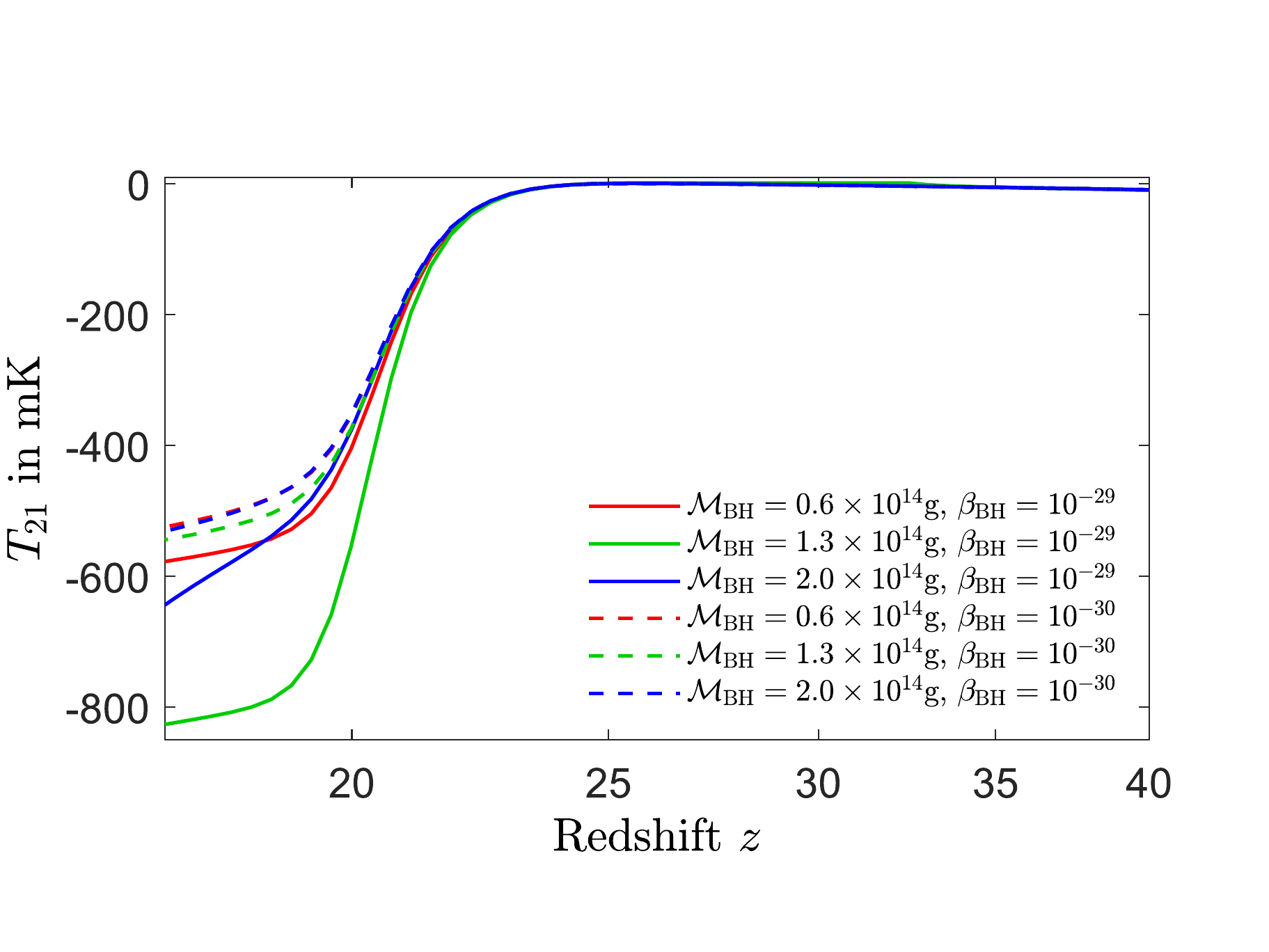}\\
			(a)&(b)\\
		\end{tabular}
		\caption{\label{fig:mchivar} Variation of $T_{21}$ (a) for different DM masses ($m_{\chi}$) and $\sigma_{41}$ with $\mathcal{M_{\rm BH}}=10^{14}$ g and $\beta_{\rm BH}=10^{-29}$, (b) for different PBH masses and $\beta_{\rm BH}$ keeping $m_{\chi}=1$ GeV and $\sigma_{41}=1$.}
	\end{figure*}
	The combined effect of PBHs evaporation and baryon-DM interaction transforms the global 21cm signal remarkably. The contributions of the PBH parameters ($\mathcal{M_{\rm BH}}$, $\beta_{\rm BH}$) and the baryon-DM interaction parameters ($m_{\chi}$, $\sigma_{41}$) in the brightness temperature are demonstrated graphically in Fig.~\ref{fig:mchivar}. In Fig.~\ref{fig:mchivar}a the variation of $T_{21}$ is shown with three different dark matter masses and two baryon-dark matter interaction cross-sections (for fixed values of $\mathcal{M}_{\rm BH}=10^{14}$ g and $\beta_{\rm BH}=10^{-29}$). In contrast, the variation due to different PBH masses and $\beta_{\rm BH}$ are shown in Fig.~\ref{fig:mchivar}b. In this particular case, the DM mass and the cross-section are kept fixed at $m_{\chi}=1$ GeV and $\sigma_{41}=1$ respectively.
	
	\begin{figure*}
		\centering{}
		\begin{tabular}{cc}
			\includegraphics[width=0.48\textwidth]{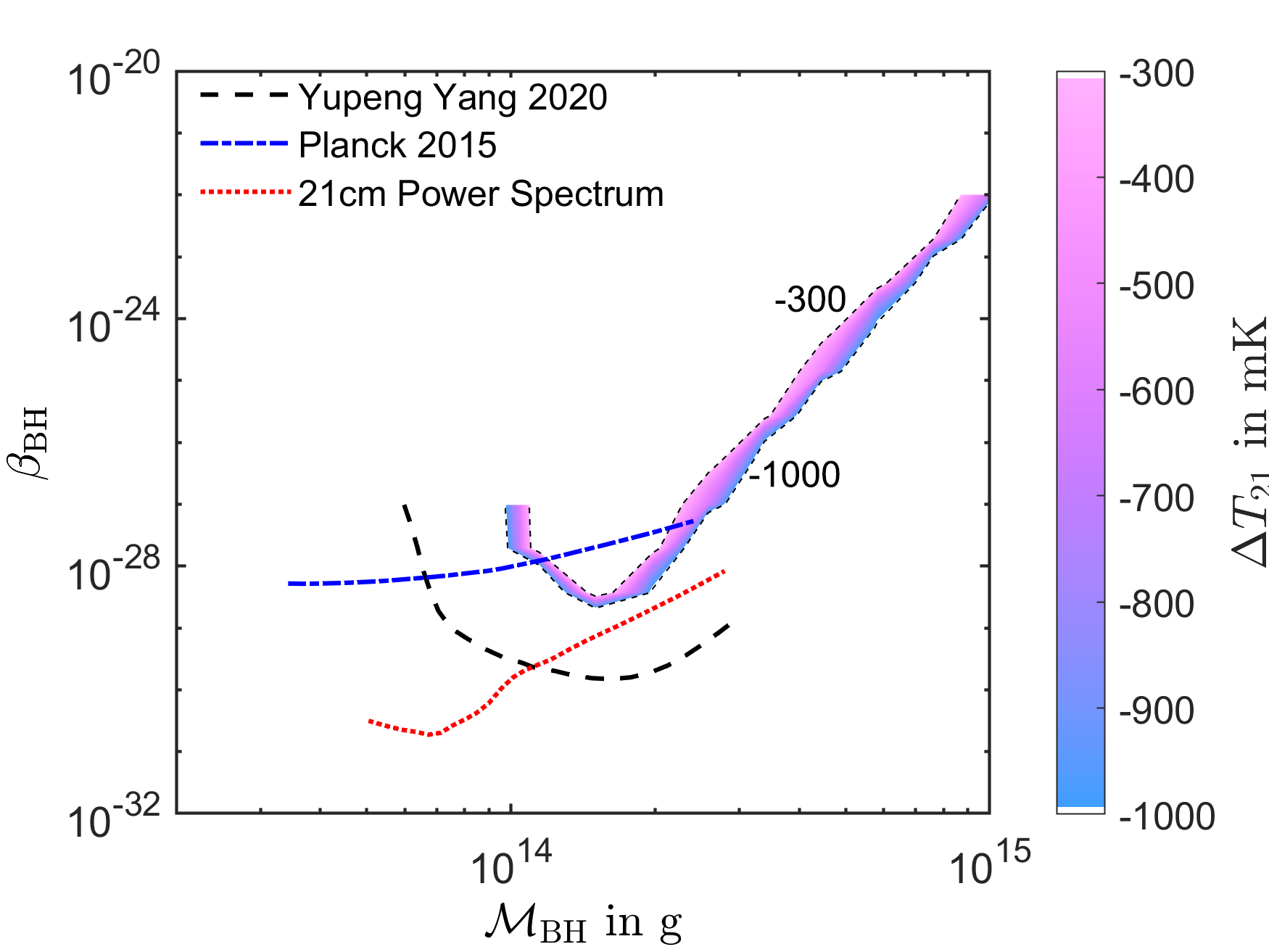}&
			\includegraphics[width=0.48\textwidth]{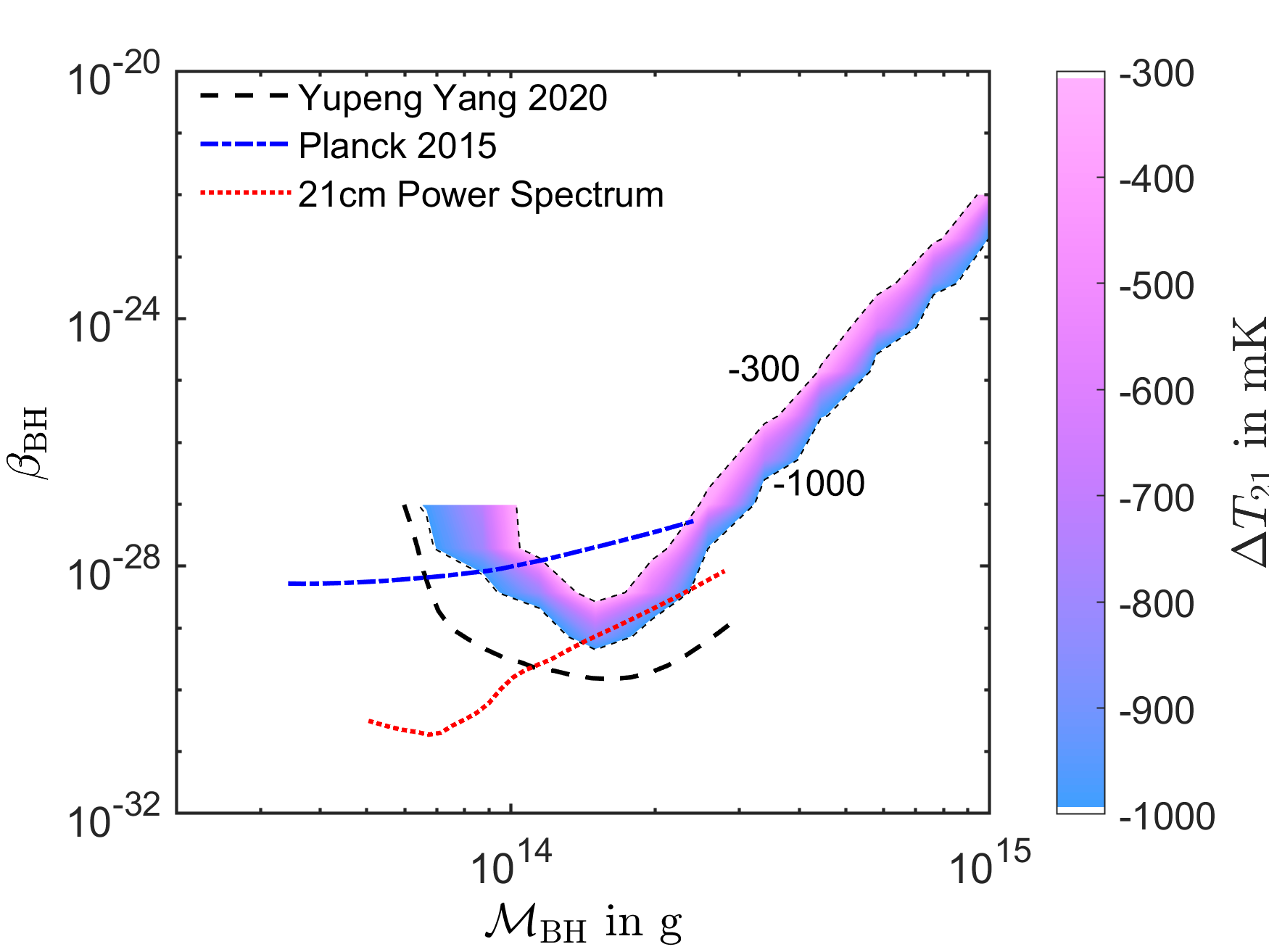}\\
			(a)&(b)\\
			\includegraphics[width=0.48\textwidth]{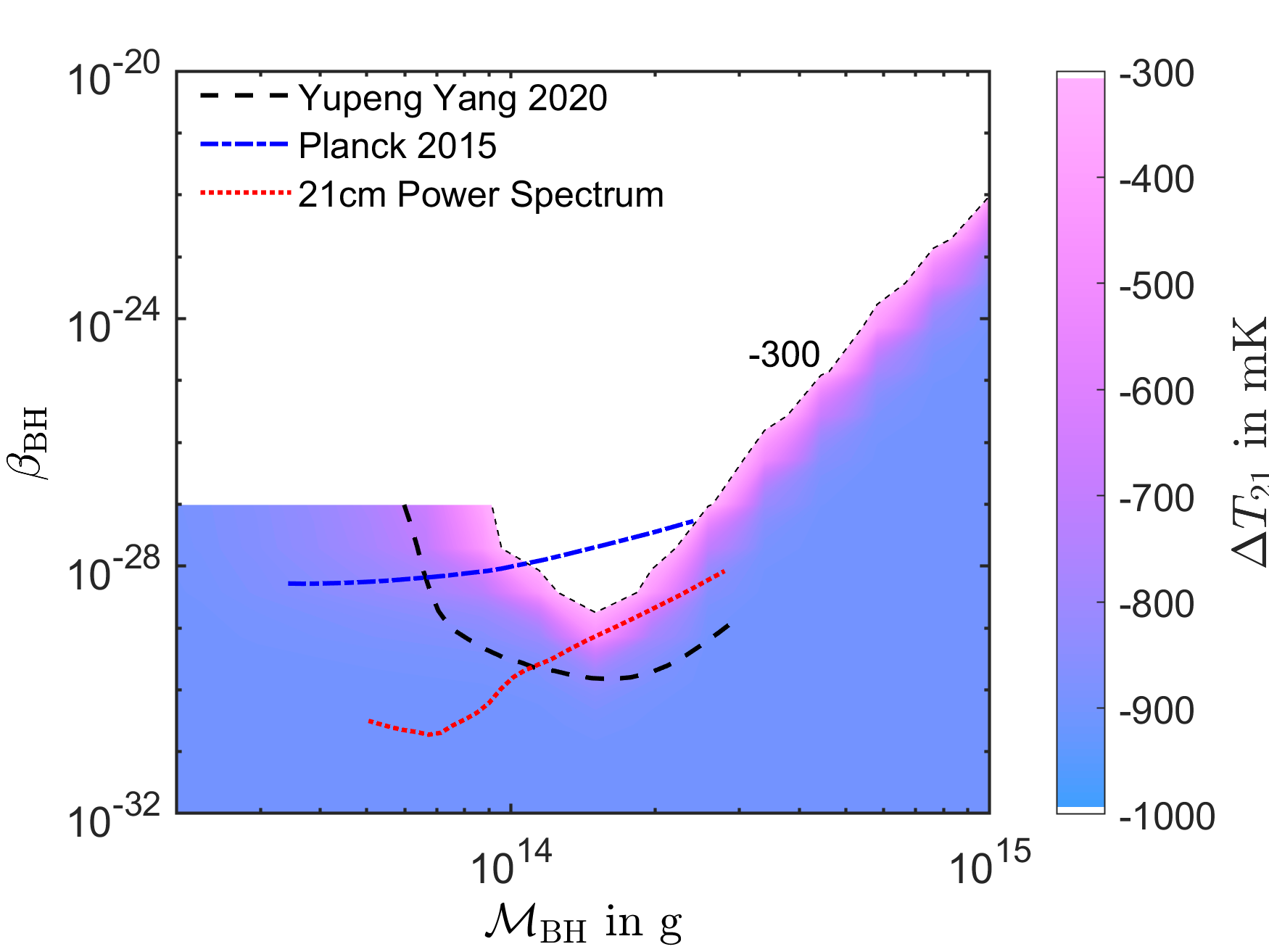}&
			\includegraphics[width=0.48\textwidth]{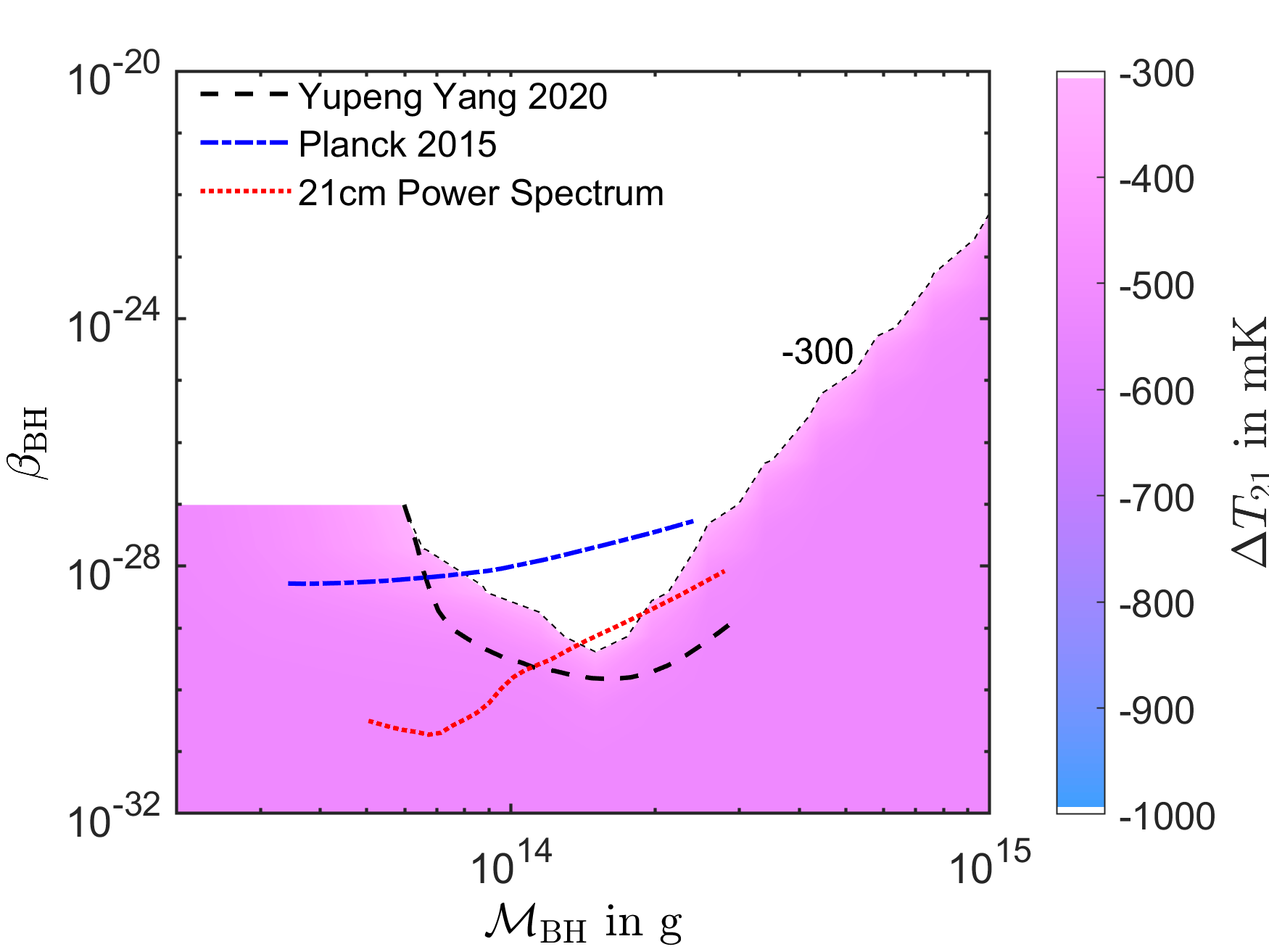}\\
			(c)&(d)\\
		\end{tabular}
		\caption{\label{fig:c_mx} The allowed zone in the $\beta_{\rm BH}$ - $\mathcal{M_{\rm BH}}$ plane for four different values of DM mass ((a) $m_{\chi}=0.1$ GeV, (b) $m_{\chi}=0.3$ GeV, (c) $m_{\chi}=0.5$ GeV and (d) $m_{\chi}=1.0$ GeV) at $\sigma_{41}=1$, that satisfy the 21cm brightness temperature limit ($-500^{+200}_{-500}$), proposed by EDGES. The black dashed line represents the upper limit for the same, as described in the work of \cite{BH_21cm_2}. The blue dash-dotted line indicates the limit obtained from the Planck 2015 data and the red dotted line shows the same that we obtained from the 21cm power spectrum \cite{BH_21cm_2}.}
	\end{figure*}
\begin{figure*}
		\centering{}
		\begin{tabular}{cc}
			\includegraphics[width=0.48\textwidth]{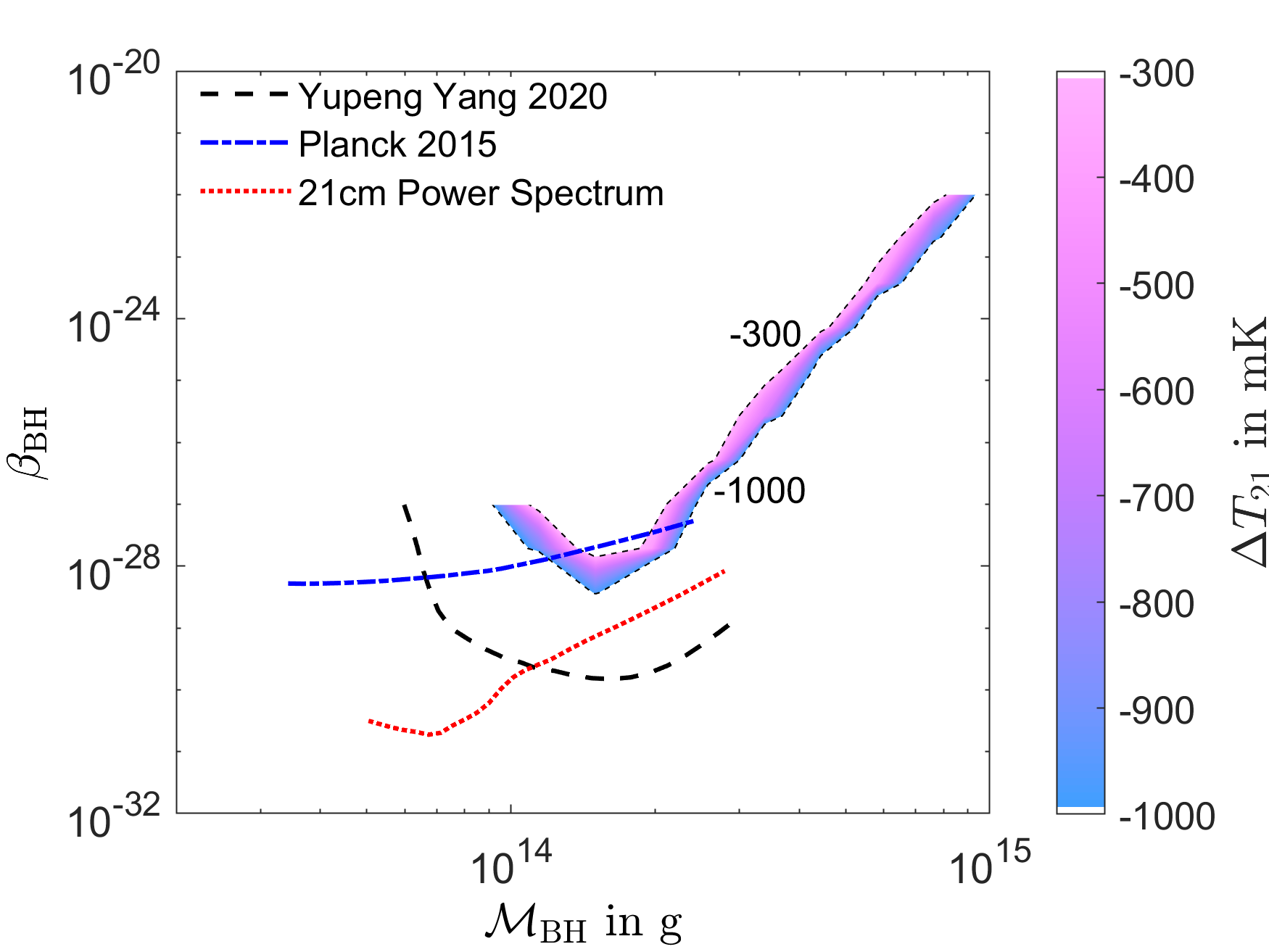}&
			\includegraphics[width=0.48\textwidth]{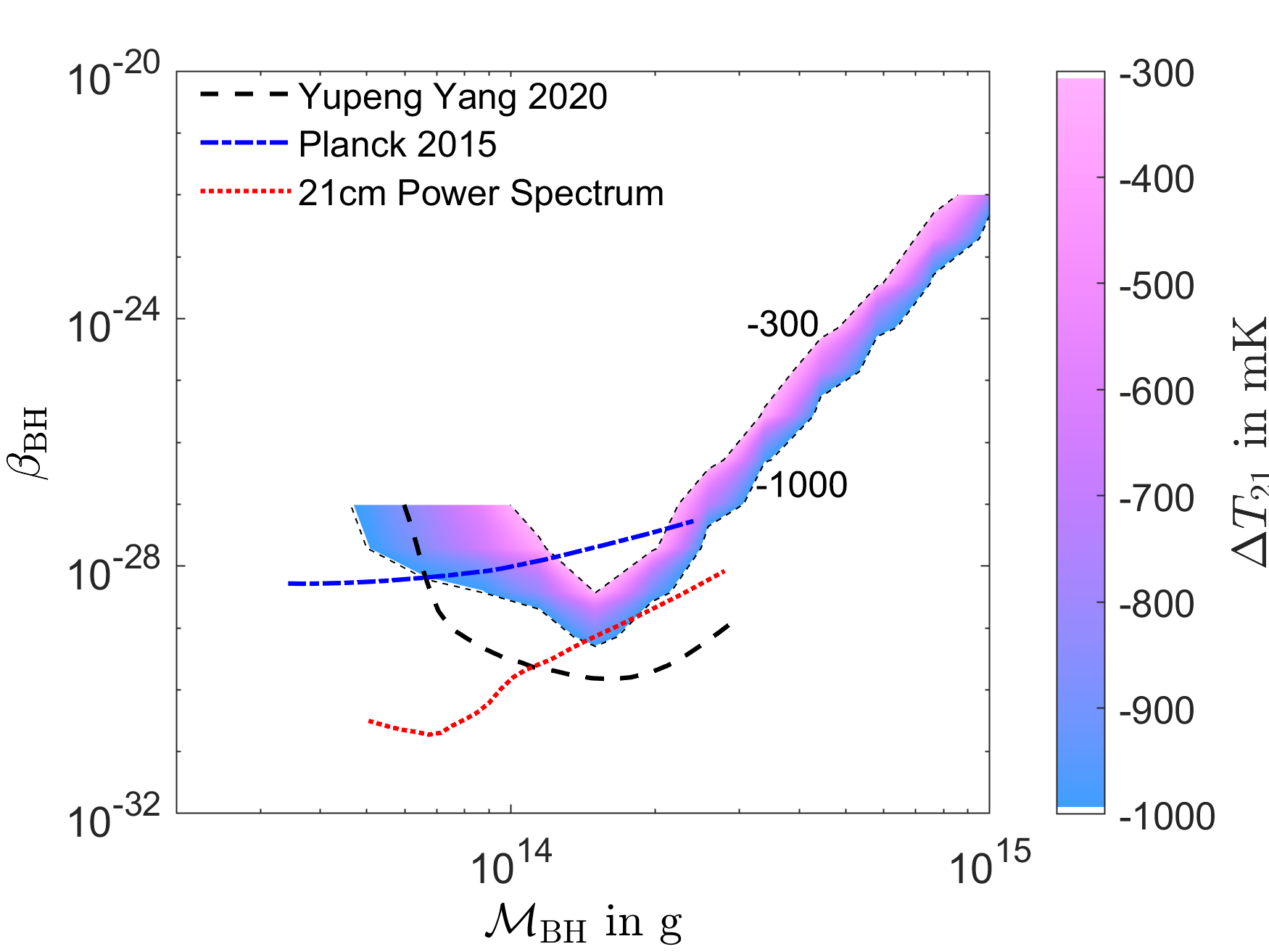}\\
			(a)&(b)\\
			\includegraphics[width=0.48\textwidth]{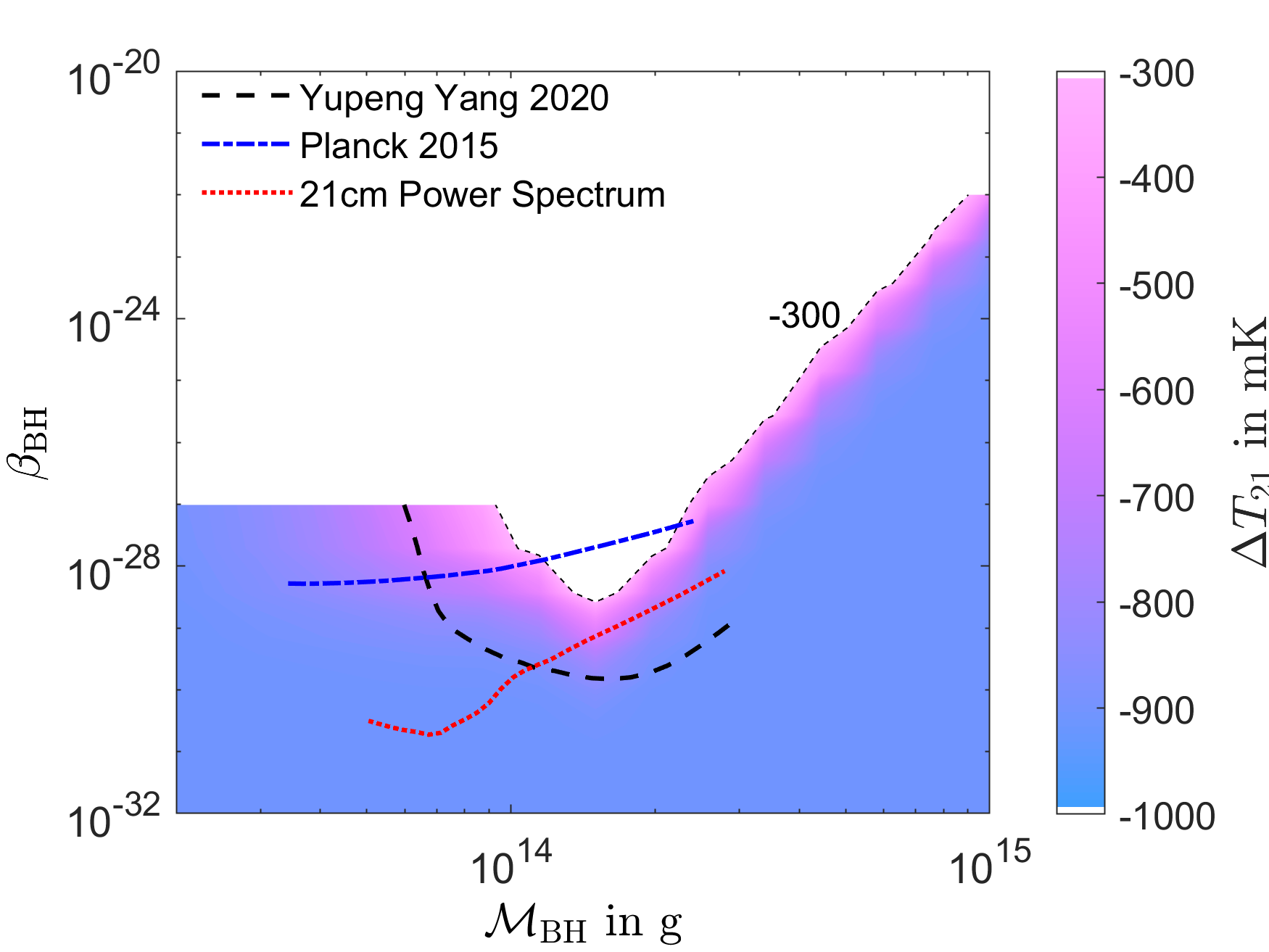}&
			\includegraphics[width=0.48\textwidth]{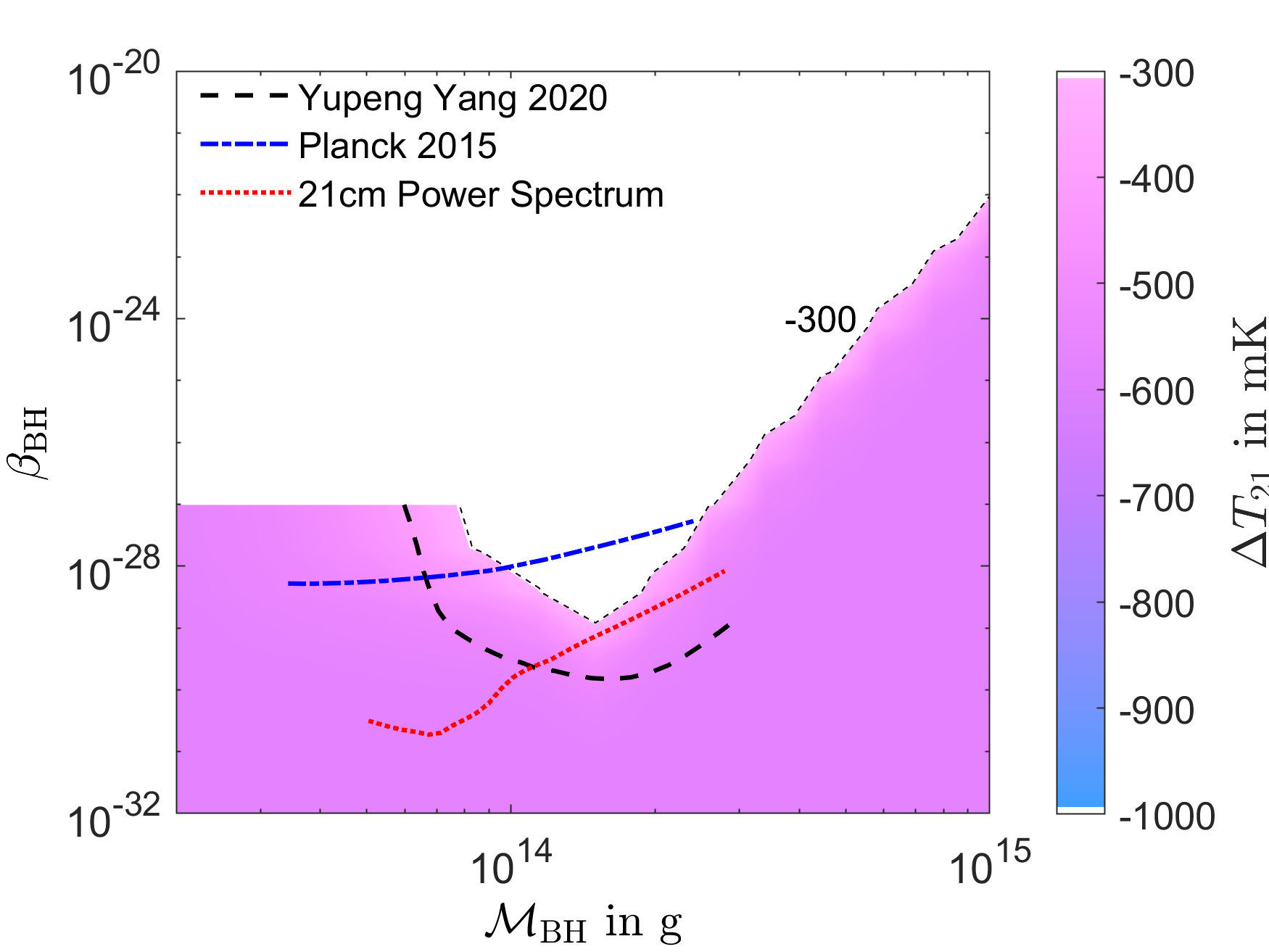}\\
			(c)&(d)\\
		\end{tabular}
		\caption{\label{fig:c_mx_sigma5}Same as Fig.~\ref{fig:c_mx} for $\sigma_{41}=5$.}
	\end{figure*}

	In the current analysis, we mainly focused on investigating the upper and lower bounds (allowed range) for the initial mass fraction of PBHs ($\beta_{\rm BH}$) for a wide range of $\mathcal{M}_{\rm BH}$ ({$6\times 10^{13}\leq\mathcal{M}_{\rm BH}\leq 10^{15}$ g}). The variations of that bound with baryon-DM interaction parameters (i.e. $m_{\chi}$ and $\sigma_{41}$) are also described in the plots of Figs.~\ref{fig:c_mx} and Fig.~\ref{fig:c_mx_sigma5}. As we estimate the limits using the consequences of the EDGES result ($T_{21} = -500^{+200}_{-500}$ at $z=17.2$), only the brightness temperature at redshift $z=17.2$ is compared with the EDGES limit. Consequently, we introduce a new parameter $\Delta T_{21}$, that represents the 21cm brightness temperature at the redshift $z=17.2$. In Fig.~\ref{fig:c_mx}, we furnish the plots describing the allowed zone in the $\mathcal{M_{\rm BH}}$-$\beta_{\rm BH}$ parameter plane for different values of DM mass $m_{\chi}$. All four plots are generated for a constant value of cross-section $\sigma_{41}=1$. Fig.~\ref{fig:c_mx}a represents the upper and lower bounds of the PBH mass fraction in the $\beta_{\rm BH}$ - $\mathcal{M_{\rm BH}}$ parameter plane for dark matter mass $m_{\chi}=0.1$ GeV, where the upper limit ($-300$ mK) and the lower limit ($-1000$ mK) are obtained from the consequences of the EDGES results ($-1000 \leq \Delta T_{21}< -300$ mK). Fig.~\ref{fig:c_mx}b, Fig.~\ref{fig:c_mx}c and Fig.~\ref{fig:c_mx}d are the similar graphs for $m_{\chi}=0.3$ GeV, 0.5 GeV and 1.0 GeV respectively. The limits obtained from our current analysis are also compared with three other limits which are described in the work of \citet{BH_21cm_2}. From Fig.~\ref{fig:c_mx}, it can be noticed that, in the case of lower $m_{\chi}$ ($0.1<m_{\chi}<0.3$ GeV), the limiting zone (the area between the upper and the lower limits) is extremely narrow. But as we increase the numerical value of $m_{chi}$, the lower bound falls abruptly keeping the upper bound almost the same (actually the upper limit decreases slightly).
	
	We also repeat the previous analysis for another value of $\sigma_{41}$ ($\sigma_{41}=5$). In this particular case, the allowed regions in the $\beta_{\rm BH}$ - $\mathcal{M_{\rm BH}}$ plane are found to be little wider than that for the $\sigma_{41}=1$ (in the case of $m_{\chi}=$0.1 GeV and 0.3 GeV). Although we investigate the upper and lower limits simultaneously, the obtained upper limit is comparable with the limit as described in the work of \citet{BH_21cm_2}. The nature of the upper bound obtained in the current work is a little analogous (but higher in amplitude) with the same as presented in the work of \citet{BH_21cm_2} at $\mathcal{M_{\rm BH}}<1.5 \times 10^{14}$ g (for $m_{\chi} =$0.5 GeV and 1 GeV). In contrast, for $m_{\chi}=0.1$ GeV and 0.3 GeV, our analysis matches with the upper bound, obtained from the Planck 2015 data at $\mathcal{M_{\rm BH}}>1 \times 10^{14}$ g (especially for the case of $\sigma_{41}=1$). Moreover, the bound from the 21cm power spectrum fits well in this range of PBH mass ($\mathcal{M_{\rm BH}}>1 \times 10^{14}$ g) in the case of higher values of $m_{\chi}$ ($m_{\chi}$=0.5 GeV and 1 GeV).
	
	\begin{figure}
		\includegraphics[width=0.7\linewidth]{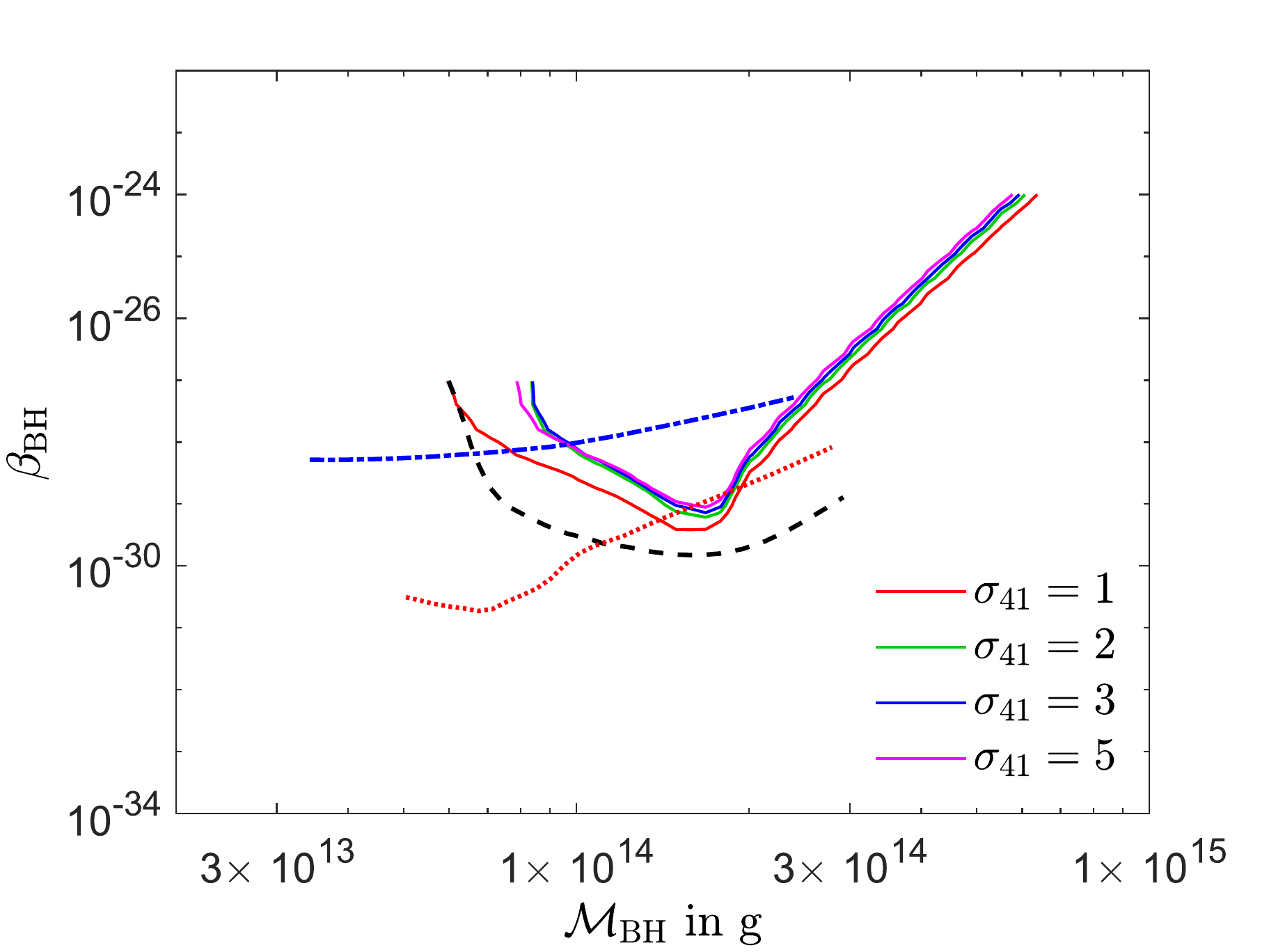}
		\caption{\label{fig:sigma_PBH} Same as Fig.~\ref{fig:c_mx} for different values of $\sigma_{41}$ keeping $m_{\chi}$ fixed at 1 GeV.}
	\end{figure}
	The variation of the allowed zone in the $\mathcal{M_{\rm BH}}$ - $\beta_{\rm BH}$ space is also evaluated for different values of baryon-DM interaction cross-section (represented by $\sigma_{41}$) keeping $m_{\chi}$ fixed at 1 GeV. In this case, only the upper bound is shown in Fig.~\ref{fig:sigma_PBH} along with the limits proposed by other works. From Fig.~\ref{fig:sigma_PBH}, it can be seen that, for a fixed value of $m_{\chi}$, the upper bound decreases with $\sigma_{41}$. The bounds corresponding to the different values of $\sigma_{41}$ fit well with the same as obtain from the 21cm power spectrum (for $\mathcal{M_{\rm BH}}>1.2\times 10^{14}$ g) and from the work of \citet{BH_21cm_2} (for $\mathcal{M_{\rm BH}}<1.2\times 10^{14}$ g). The nature of the bound obtained in the work of \citet{BH_21cm_2} is also similar for smaller values of $\sigma_{41}$ ($\sigma_{41}\approx 1$).
	
	\begin{figure}
		\includegraphics[width=0.7\linewidth]{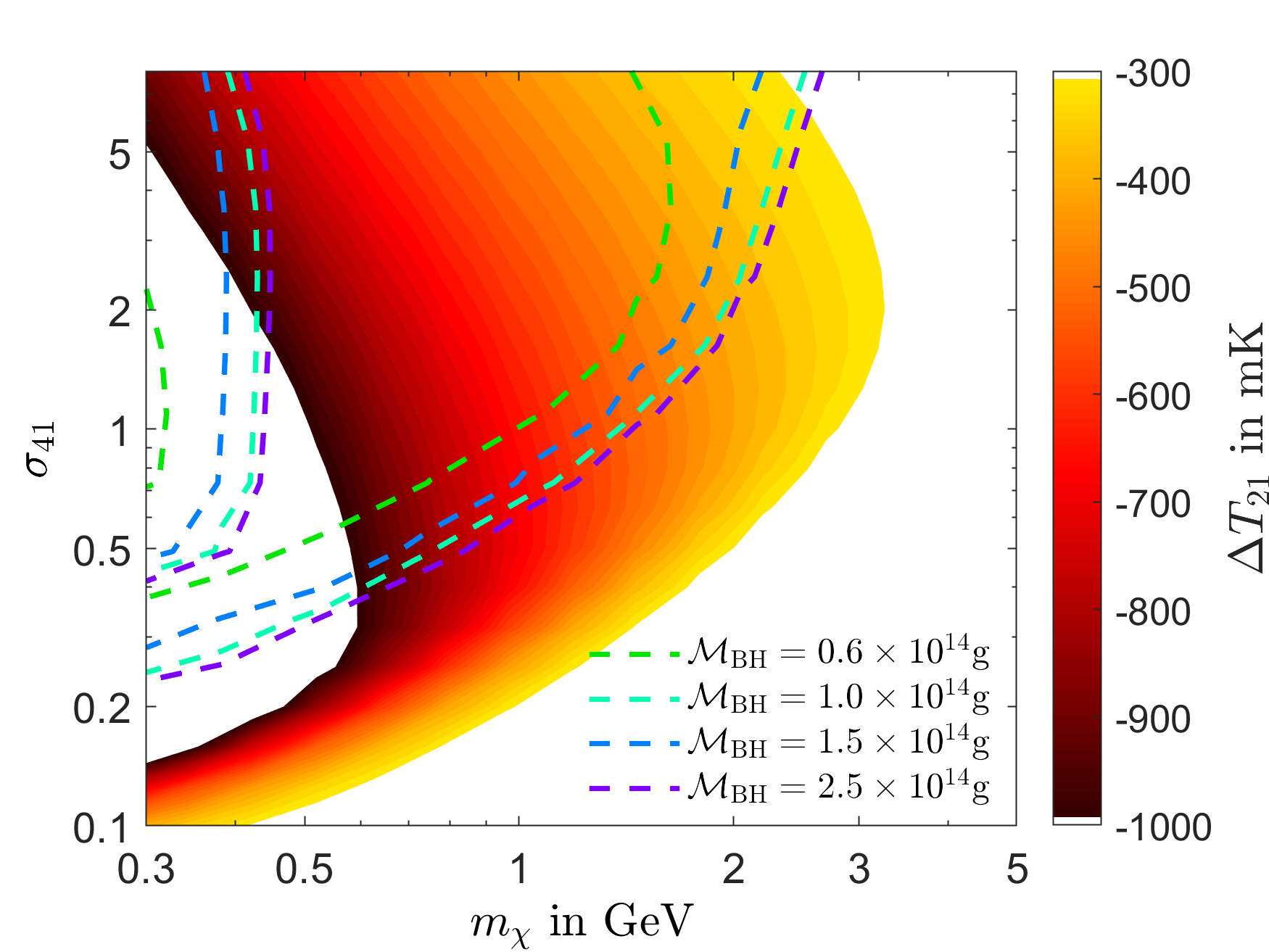}
		\caption{\label{fig:mchi_sigma} The allowed region of $m_{\chi}$  and $\sigma_{41}$. The color region represents the allowed region in the case, where the presence of PBH is ignored. The different colored dashed lines are addressing the boundaries of the allowed zone where the effect of the PBHs of different masses is taken into the account. The chosen values of $\beta_{\rm BH}$ for each case are described in the text.}
	\end{figure}
	Eventually, we address similar bounds in the $m_{\chi}$ - $\sigma_{41}$ space for different values of PBH masses ($\mathcal{M_{\rm BH}}$). In Fig.~\ref{fig:mchi_sigma}, the allowed range for the dark matter mass $m_{\chi}$ and $\sigma_{41}$ is demonstrated (colored region) in absence of the effect of Hawking radiation. In this particular case, the different colors correspond to the different values of $\Delta T_{21}$ in mK (see colorbar). However the allowed zone is modified as the heating in the form of Hawking radiation is taken into account. In this figure (Fig.~\ref{fig:mchi_sigma}) the same allowed zones are addressed where different sets of PBH parameters are considered in the system namely ($\mathcal{M_{\rm BH}}=0.6\times 10^{14}$ g, $\beta_{\rm BH}=1.0\times 10^{-29}$), ($\mathcal{M_{\rm BH}}=1.0\times 10^{14}$ g, $\beta_{\rm BH}=3.0\times 10^{-30}$), ($\mathcal{M_{\rm BH}}=1.5\times 10^{14}$ g, $\beta_{\rm BH}=1.5\times 10^{-30}$) and ($\mathcal{M_{\rm BH}}=2.5\times 10^{14}$ g, $\beta_{\rm BH}=5.3\times 10^{-28}$). In this figure (Fig.~\ref{fig:mchi_sigma}) it appears that the allowed region in the $m_{\chi}$-$\sigma_{41}$ plane shifts toward the lower values of DM mass $m_{\chi}$ but higher in $\sigma_{41}$ while the heating due to the PBH is incorporated into the system. It is to be mentioned that, the limits of the $m_{\chi}$ for the individual cases agree with the outcome of the work of \citet{rennan_3GeV} (i.e. $m_{\chi} \leq 3$ GeV).
	
\section{\label{sec:conc} Summary and Discussions}
	In this work we have studied the limits of the initial PBH mass fraction ($\beta_{\rm BH}$), the initial mass of PBH ($\mathcal{M_{\rm BH}}$), DM mass ($m_{\chi}$) and $\sigma_{41}$ in the framework of the 21cm cosmology. The mutual contribution of baryon-DM interaction and PBH in the ISM heating modifies the 21cm brightness temperature significantly and hence the bounds of the initial mass fraction of primordial black holes. In the current analysis, we evolve five coupled equations (Eq.~\ref{eq:PBH}, Eq.~\ref{eq:T_chi}, Eq.~\ref{eq:T_b}, Eq.~\ref{eq:V_chib} and Eq.~\ref{eq:xe}) simultaneously with cosmological redshift $z$ in order to estimate the temperature evolution of the IGM. In Fig.~\ref{fig:tspin} the variations of baryon temperature as well as the spin temperature are shown with different chosen values of DM mass, baryon-DM cross-section and the PBH masses. The changes in the brightness temperature are also pictured in the two different plots of Fig.~\ref{fig:mchivar}. 
	
	We put forward the allowed zone in the parameter plane of the initial mass fraction of PBHs and $\mathcal{M_{\rm BH}}$ for different chosen values of $m_{\chi}$ and $\sigma_{41}$. In this analysis, the allowed limits are estimated by incorporating the EDGES's limit on the 21cm brightness temperature. In Fig.~\ref{fig:mchi_sigma}, it can be seen that the allowed upper and lower bounds are found very closed when the smaller values of $m_{\chi}$ are considered. However as the numerical value of $m_{\chi}$ increases, the lower bound falls abruptly and almost disappears at $m_{\chi}\gtrapprox0.5$ GeV. The identical variation is also investigated for $\sigma_{41}=5$ and compared with the limits addressed in other works \cite{BH_21cm_2,BH_21cm_3}. It is to be mentioned that, the estimated upper limits fit well with the limit as obtained from the Planck data (2015), when the lower masses of DM are considered ($m_{\chi}\approx 0.1$ GeV). The obtained bounds are also comparable to the same as proposed by \cite{BH_21cm_2} in the case of higher $m_{\chi}$. Moreover, the limits from the 21cm power spectra fits well for $\mathcal{M}_{\rm BH}\geq 1.1 \times 10^{14}$ g and $m_{\chi} \approx 1$ GeV. The similar bound for $m_{\chi}$ and $\sigma_{41}$ is also described in Fig.~\ref{fig:sigma_PBH}, where the upper bound of $m_{\chi}$ agree with the work of \citet{rennan_3GeV} for each value of $\mathcal{M_{\rm BH}}$ considered in the calculation. Hopefully, future investigation in 21cm physics will enrich our understanding with several unexplored aspects of the dark Universe \cite{Burns_2017,plice2017dare}.

\section*{Acknowledgements}
	Two of the authors (A.H. and S.B.) wish to acknowledge the support received from St. Xavier's College. A.H. also acknowledges the University Grant Commission (UGC) of the Government of India, for providing financial support, in the form of NET-SRF.	

\bibliography{PUB21}

\end{document}